\documentclass[12pt]{article}
\usepackage{epsf}
\usepackage{graphicx}
\usepackage{lscape}
\usepackage{amssymb}
\usepackage{pazh_eng}
\usepackage{sidecap}
\usepackage{float}
\usepackage{pstricks}

\tightenlines

\begin{document}

\title{\bf Search for Outbursts in the Narrow 511-keV Line from Compact Sources
Based on INTEGRAL Data}

\author{\bf \hspace{-1.3cm}\copyright\, 2010  \ \
S.S.Tsygankov\affilmark{1,2}$^{\,*}$ and E.M.Churazov\affilmark{1,2}}
 \affil{$^1$ {\it Max Planck Institut f\"ur Astrophysik, Karl-Schwarzschild-Str. 1, Postfach 1317, D-85741 Garching, Germany}\\
$^2$ {\it Space Research Institute, Russian Academy of Sciences,
   Profsoyuznaya ul. 84/32, Moscow 117997, Russia}}

\vspace{2mm}

Received September 14, 2009

\sloppypar
\vspace{2mm}
\noindent


We present the results of a systematic search for outbursts in the narrow positron annihilation
line on various time scales ($5\times10^{4}$~ -- ~$10^{6}$~ s) based
on the SPI/INTEGRAL data obtained from 2003 to 2008.
We show that no outbursts were detected with a statistical
significance higher than $\sim$$6\sigma$ for any of the time
scales considered over the entire period of observations. We also show that, given the large number of
independent trials, all of the observed spikes could be associated with purely statistical flux fluctuations
and, in part, with a small systematic prediction error of the telescope's instrumental background. Based on
the exposure achieved in $\sim$6 yr of INTEGRAL operation, we provide conservative upper limits on the rate
of outbursts with a given duration and flux in different parts of the
sky.


\vfill

{$^{*}$ E-mail: tsygankov@iki.rssi.ru}
\newpage
\thispagestyle{empty}
\setcounter{page}{1}

\section*{INTRODUCTION}

The positron annihilation line has been studied in
detail by many authors based on data from various experiments
since it was first discovered at the Galactic
center (Johnson et al. 1972). Several positron generation
models that completely or partly describe the
observed line flux have been proposed to explain the
nature of the observed emission: the synthesis of radioactive
elements with the $\beta^{+}$
 decay channel among
daughter nuclei ($^{22}Na$,~ $^{26}Al$,~ $^{44}Ti$,~
$^{56}Ni$,~ $^{57}Ni$) through
supernova explosions, gamma-ray bursts, and hypernovae;
the production of electron-positron pairs near
pulsars and black holes; the interaction of cosmic rays
with interstellar matter and the annihilation of dark matter
particles (for a review, see, e.g., Teegarden and
Watanabe 2006; Bandyopadhyay et al. 2009). In all of
the positron generation mechanisms considered, the
initial kinetic energy of the positrons is comparable to
or appreciably higher than the electron rest mass.

If compact stellar objects are assumed to be
the sources of positrons (see, e.g., Prantzos 2004;
Weidenspointner et al. 2008), then the observed
annihilation radiation need not be constant and can
come in the form of outbursts. The slowing-down and
annihilation time scale for positrons with a typical
MeV energy is determined mainly by the density of
the medium ($n$): 
$\displaystyle\tau \sim 10^5 \left(\frac{n}{10^7 {\rm
    cm^{-3}}}\right)^{-1}$~ s 
(Forman et al. 1986). Near compact sources (e.g., in accretion
disks or on a stellar surface), the density can be very
high and the actual constraint on the lifetime of a
positron (including its slowdown and annihilation)
can be related to the positron time of flight, of the
order of seconds or shorter.

Possible flux variability at the Galactic center near
0.5 MeV was pointed out by a number of authors
(see, e.g., Riegler et al. 1981). However, it was subsequently
shown that the 511-keV line flux measured
by a specific instrument clearly depended on the size
of the instrument's field of view; the larger the telescope's
field of view, the higher the flux it recorded
(see, e.g., Teegarden 1994). In this way, a substantial
fraction of the reports on flux variability in the annihilation
line were disproved. It was also concluded
from the detected dependence that the source of the
annihilation radiation at the Galactic center is diffuse
in nature.

In addition to the Galactic center, evidence for
possible outbursts near 0.5 MeV has been found
for compact sources: 1E 1740.7-2942 (Bouchet
et al. 1991; Sunyaev et al. 1991) and Nova Musca
(Sunyaev et al. 1992; Goldwurm et al. 1992, Gilfanov
et al. 1994). The recorded broad emission lines in their
spectra were considerably shifted relative to 511 keV
to lower energies and were observed on a time scale
of $\sim$0.5-2 days.

There are also a number of papers in which it is
argued that the positron annihilation line flux from
the Galactic-center region is constant (for a review,
see., e.g., Purcell et al. 1997). The results of a search
for transient events in the data of the first year of
INTEGRAL operation are presented in Teegarden
and Watanabe (2006) and also reveal no significant
variability. Below, we present the results of a systematic
search for outbursts in a narrow energy band,
508-514 keV, near the electron-positron annihilation
line (about 99\% of the flux in the annihilation
line recorded from the Galactic center and a negligible
flux from the continuum of sources with a
hard energy spectrum fall within this band) based
on six-year-long all-sky SPI/INTEGRAL observations.
Our analysis was performed on outburst time
scales from $5\times10^4$~ to $10^6$~ s.

\section*{OBSERVATIONS}

Our work is based on data from the SPI spectrometer
(Vedrenne et al. 2003) onboard the INTEGRAL
observatory (Winkler et al. 2003). This is one of the
main INTEGRAL instruments and operates in the
energy range 20--8000 keV. Owing to its high energy
resolution ($\sim$2 keV at energy 511 keV), it is ideally
suited to investigating the properties of the positron
annihilation line in the Galaxy. The spectrometer consists
of 19 individual cooled germanium detectors
(17 detectors remained operational after July 2004).

The main goal of this paper is a ``blind'' search
(when the position of the sought-for object is not
known in advance) for variable emission sources in
the positron annihilation line. With such a formulation
of the problem, the simplest and most reliable
outburst detection method is to search for features on
the count rate curve of the telescope. In this approach,
the telescope's optical scheme is roughly equivalent
to a collimator with a field of view $\sim$30\deg~ in diameter.
Using a coded aperture and image reconstruction
gives a gain in sensitivity only if the position of the
outburst source is known in advance.

The INTEGRAL observatory has a high-apogee
orbit with a revolution period around the Earth of
about 3 days. The orbital parameters were chosen in
such a way that the satellite is located outside the
Earth's radiation belts for $\sim$90\% of the time, which
allows continuous observations to be performed by
the INTEGRAL instruments for a long time. The
individual observations are the pointings of the INTEGRAL
axis toward specific regions of the sky and
have a typical duration of $\sim$2000 s. We used all of
the available INTEGRAL data from February 2003 to
October 2008 (the total exposure time was $\sim$$7\times10^{7}$~ s).

{\centerline{}}
{\centerline{\it Outburst Search Algorithms}}

For our analysis, we divided the entire celestial
sphere into pixels 4$\times$4 deg. in size, which roughly
corresponds to the SPI angular resolution. A light
curve consisting of the fluxes in an individual pointing
in the 508-514 keV energy band was constructed in
each of these pixels. Since the SPI field of view is
much larger than the size of a single map pixel, each
INTEGRAL pointing makes a contribution immediately
to several tens of pixels on the map. Accordingly,
the effective area of the spectrometer for each of these
map pixels will be not the same but distributed in
accordance with the shape of the SPI response at the
energy under consideration (Sturner et al. 2003). The
photon flux for each pointing in a specific map pixel
was defined as the count rate averaged over all active
detectors in the 508-514 keV energy band divided
by the effective area of the spectrometer at energy
$\sim$508 keV assigned to this map pixel in this pointing.

\begin{figure*}[t]
\centerline{\includegraphics[width=14cm,bb=20 120 540 380,clip]{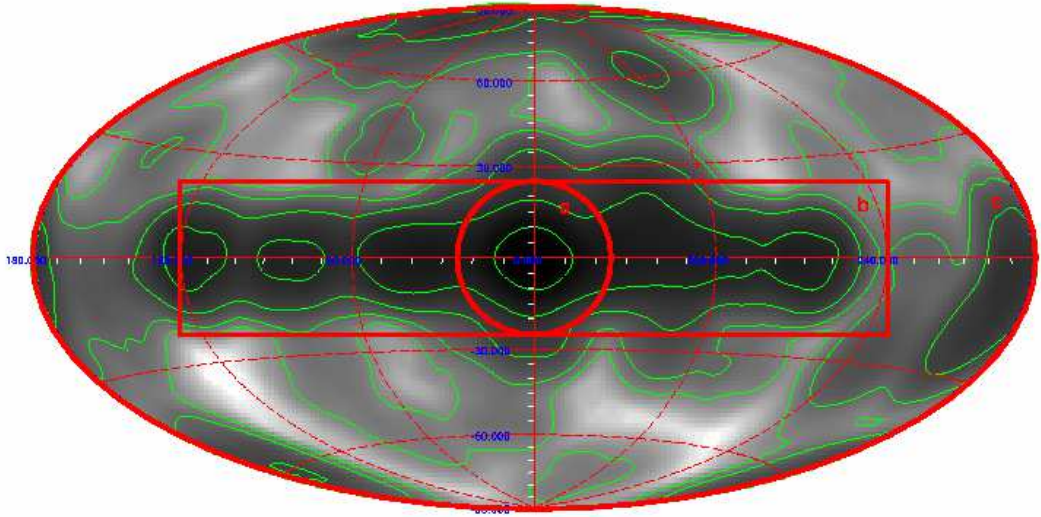}}

\vfill

\caption{Exposure map of the SPI observations used here. The heavy solid lines indicate the three regions for which the results
of our analysis are presented (a -- the Galactic center (the
distance from the center is R$<$25\deg), b -- the Galactic plane
(-120\deg$<l<$120\deg, -25\deg $<b<$25\deg), c -- the entire sky); 
the contours represent the levels of equal exposure 0.07, 0.2, 0.7,
2, and 6 Ms.}\label{expo}
\end{figure*}


Since the distribution of the effective area over the
SPI field of view is highly nonuniform, the presence
of persistent annihilation emission sources in the sky
must lead to significant variations in the measured
flux with source position in the SPI field of view. To
eliminate these effects related to the presence of a
persistent bright emission source in the 511-keV line
at the Galactic center, its contribution was subtracted
by assuming a bivariate Gaussian intensity distribution
of this source with the center at the Galactic
center and with a full width at half maximum
(FWHM) along and across the Galactic plane of 9\deg
and 6\deg, respectively (the parameters were taken from
Churazov et al. (2010)).

Thus, the flux in units of phot cm$^{-2}$ s$^{-1}$ 
for a
given pointing $i$ in a map pixel with coordinates $(x,y)$
was calculated as

\begin{equation}\label{eq1}
F_i(x,y)=\frac{I_{det,i}(x,y)-I_{bkg,i}(x,y)-\int GC_{mod}(\alpha,\delta)\times A_{eff}(\alpha,\delta,x,y) \mathrm{d}\alpha \mathrm{d}\delta}{A_{eff}(x,y)},
\end{equation}

where $I_{det,i}(x,y)$ and $I_{bkg,i}(x,y)$
 are the detected and
background (see below) count rates averaged over
all active detectors for pointing $i$ in map pixel $(x,y)$,
respectively; $A_{eff}(\alpha,\delta,x,y)$
 is the mean response of
the SPI detectors in map pixel $(x,y)$ to a source of
unit intensity with celestial coordinates $(\alpha,\delta)$. 
For a more accurate determination of the contribution
from the persistent source at the Galactic center, the
integral in Eq. (1) was taken on a 0\deg.5$\times$0\deg.5 grid.

In each map pixel, the entire observing time was
divided into intervals of duration $\tau$ 
 (from $5\times10^4$~ to $10^6$~ s). Next, the flux within each interval $j$
was averaged as 
$<F_{(x,y),j}>=\frac{\Sigma (F_{(x,y),i}/\sigma_{(x,y),i}^2)}{\Sigma (1/\sigma_{(x,y),i}^2)}$,
 where $<F_{(x,y),j}>$ 
is the weighted mean flux in interval $j$,
$F_{(x,y),i}$ and $\sigma_{(x,y),i}$
 are the flux and the statistical
flux error in the 508-514 keV channel in pointing $i$
in map pixel $(x,y)$, respectively. Subsequently, we
analyzed the deviation of this quantity from the mean
flux $<F_{(x,y)}>$
 over the entire light curve in a given map
pixel obtained in a similar way. The significance of
the flux deviation from the mean was calculated as
$\frac{<F_{(x,y),j}>-<F_{(x,y)}>}{\sigma_{<F_{(x,y),j}>}}$. 
We searched for outbursts in the
511 keV line on four time scales: 50, 100, 500, and
1000 ks.

As a result of the algorithm described above, we
obtained the maps for each averaging time scale in
each pixel of which the highest significance of a positive
spike in flux relative to the mean in this map pixel
was reflected.

{\centerline{}}
{\centerline{\it Data Selection}}

Before their use, the data were cleaned from the
observations performed during or immediately after
powerful solar flares, detector calibration periods, and
other events unrelated to the sought-for astrophysical
effects.

Due to peculiarities of the program of INTEGRAL
observations, in our analysis we divided the entire celestial
sphere into three parts (the Galactic center, the
Galactic plane, and the entire sky) differing greatly
in typical exposure time. These regions and the total
exposure map are shown in Fig. 1. The exposure time
in each map pixel was defined as the total exposure
time $T_{i}$ in a given observation $i$ corrected for the ratio
of the instrument's effective area $A_{eff}(x,y)_{i}$
 calculated
for the telescope's orientation in observation $i$ and
coordinates on the celestial sphere corresponding to
pixel $(x,y)$ to the maximum efficiency over the field
of view at energy $\sim$508 keV ($max (A_{eff})$). Thus, the
formula used to calculate the exposure time in each
map pixel appears as

$T(x,y)=\frac {\sum^{N}_{i=1} T_{i} \times A_{eff}(x,y)_{i}} {max (A_{eff})} $,

where $i$ is the pointing number, $N$ is the number of
INTEGRAL pointings covering map pixel $(x,y)$, $T_{i}$ is
the total exposure time of observation $i$, $max (A_{eff})$ is
the maximum effective area of the spectrometer over
the field of view at energy $\sim$508 keV, $A_{eff}(x,y)_{i}$
is the
effective area of the spectrometer in map pixel $(x,y)$
for pointing $i$.

\begin{figure*}[t]
\centerline{\includegraphics[width=10cm,bb=20 275 515 690,clip]{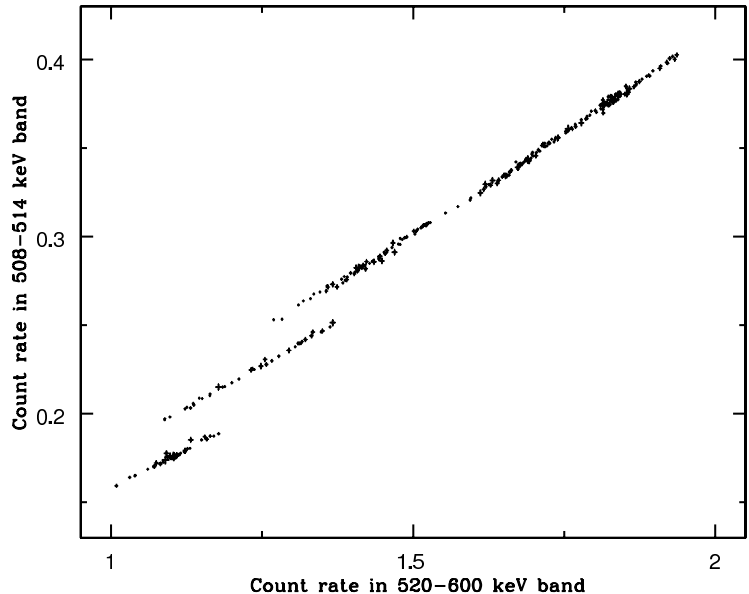}}

\vfill

\caption{Correlation of the count rates in the 508-514 and 520-600 keV energy bands. Each point corresponds to the count
rate averaged over the observations within one INTEGRAL revolution. We used only the pointings during which the Galactic
plane and bright sources detected in hard X-rays were not observed. The abrupt changes in the correlation of the count rates
correspond to the successive failure of two (of the 19) SPI detectors.}\label{flcor}
\end{figure*}

We see that most of the observing time is allocated
for the Galactic-center and Galactic-plane observations.
In particular, the source 1E 1740.7-2942
whose spectrum previously exhibited a broad feature
at energies 400-500 keV (Bouchet et al. 1991; Sunyaev
et al. 1991) is located in this region.

{\centerline{}}
{\centerline{\it Background Modeling}}

The recorded flux in the 511-keV line is strongly
dominated by a variable (in time) background that
is mainly related to the flux of charged cosmic-ray
particles and their interaction with the SPI detectors
and INTEGRAL constructions. We used two different
methods to model the background flux at each
instant of time:

(1) based on the use of the detector count rate
above an upper energy threshold of 8 MeV (saturated
events rate) as an indicator of the particle
count rate (see, e.g.,Knodlseder et al. 2005; Churazov
et al. 2005);

(2) based on the correlation of the fluxes in the
508-514 and 520-600 keV energy bands.

Both methods yield similar results, but here we
used the background construction method (2) as a
more appropriate one for the problem under consideration
(see below). Since the SPI spectrometer has
a good energy resolution, the 520-600 keV channel
contains no contribution from the narrow 511-keV
line and the ortho-positron continuum. The background
count rate in the 520-600 keV energy band
exceeds the count rate in the 508-514 keV band by a
factor of 5, which allows no additional contribution to
the statistical error of our results to be made. A serious
advantage of the background model construction
based on the count rate in a channel close in energy to
the 508-514 keV one is an effective subtraction of the
systematic features that manifest themselves in both
bands.

Figure 2 shows the correlation of the count rates
in the 508-514 and 520-600 keV energy bands.
The abrupt changes in the correlation of the count
rates correspond to the successive failure of two SPI
detectors in December 2003 and July 2004. Because
of the long-period background variations, we selected
14 epochs -- the time intervals for which the correlation
coefficients between the count rates in the investigated
and reference energy channels were calculated
independently.

\begin{figure*}[t]
\centerline{\includegraphics[width=10cm,clip]{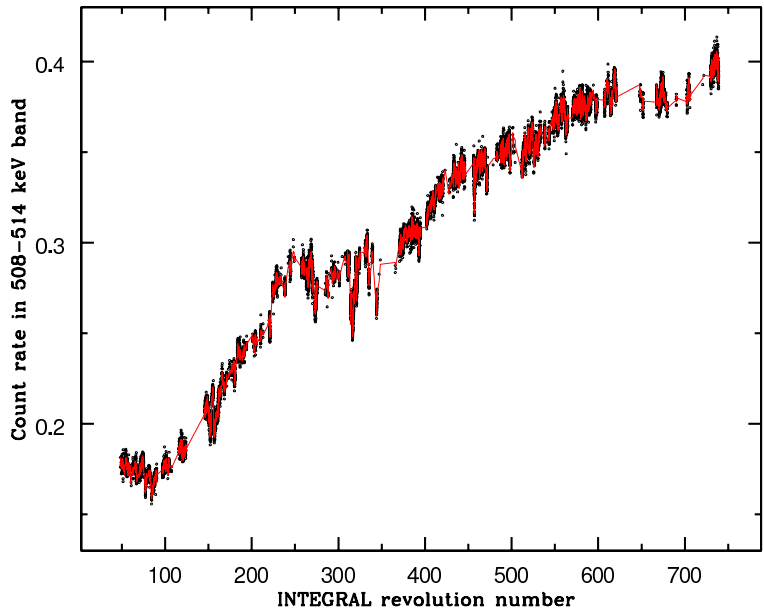}}

\vfill

\caption{Count rate in the 508-514 keV energy band versus time (in units of the INTEGRAL revolution period). Each point
corresponds to the count rate averaged over all active detectors
during one revolution. The line indicates the predicted background in the same energy
band based on model 2 (see the text).}\label{back}
\end{figure*}

It should be mentioned that the flux variations in
the 520-600 keV band related to real astrophysical
sources can cause flux variability in the 508-514 keV
band. By definition, the background subtraction
procedure is calibrated in such a way that if the ratio
of the fluxes in the 508-514 and 520-600 keV bands
is the same as that in the telescope's instrumental
background ($R_{background}=F_{508-514}/F_{520-600}\sim 0.2$), 
then the expected flux in the 508-514 keV band
after the background subtraction is zero. For typical
astrophysical sources, the flux ratio is appreciably
smaller. For example, for the spectrum of the Crab
Nebula, $R_{Crab}=F_{508-514}/F_{520-600}\sim 0.09$. Thus, a
slight increase in flux in the 508-514 keV band
will be more than compensated for by an increase
in flux in the 520-600 keV. The measured flux in
the 508-514 keV band corrected for the background
will be $F_{508-514, Crab}- R_{background}\times F_{520-600,Crab}\sim
-1.4\times10^{-4} $ phot cm$^{-2}$ s$^{-1}$. Thus, negative spikes
can appear on the light curve in the 508-514 keV
band, provided that the continuum intensity of the
source is very high. In this paper, we restrict our
analysis to searching for only the positive spikes in
the narrow 511-keV line.

The quality of the prediction of the background
radiation made by the method described above is illustrated
by Fig. 3. The points in this figure indicate
the measured count rate in the 508-514 keV energy
band averaged over all active detectors on the scale
of one INTEGRAL revolution; the solid line indicates
the predicted background.

{\centerline{}}
{\centerline{\it The Systematic Error}}

In addition to the statistical errors, there exist a
number of sources of systematic errors when working
with the SPI data. Thus, for example, the shape of the
spectrum and the normalization of the background
radiation cannot be predicted absolutely accurately.
Nonideal subtraction of persistent emission sources
in the 508-514 keV energy band can be yet another
source of systematic errors.

The total error can be estimated from the distribution
of flux significances over the entire map after
the subtraction of the background flux and the constant
flux in the 511-keV line from the source at the
Galactic center (see Eq. (1)). If the systematic errors
are small compared to the typical statistical errors in
a given series of measurements and if there are no
strong outbursts in the 511-keV line, then this must
be a Gaussian distribution with a mean of 0 and a
variance of 1.

\begin{figure*}[t]
\centerline{\includegraphics[width=10cm,bb=70 250 520 690,clip]{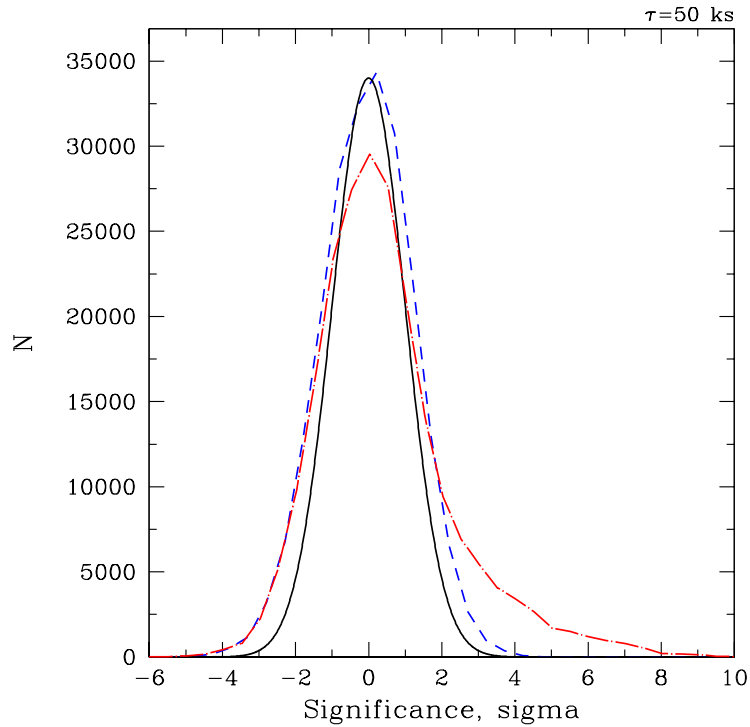}}

\vfill

\caption{Distribution of 508-514-keV flux significances over the sky before (dash-dotted line) and after (dashed line) the
subtraction of the contribution from the central persistent source for the averaging time scale $\tau=50$ ks and the second set of
errors (see the text). For comparison, the solid line indicates a Gaussian distribution with a mean of 0 and a variance of 1. The
statistical significance in units of the standard deviation is along the horizontal axis. The typical value of one standard deviation
for this time scale is $\sim2\times10^{-4}$
phot cm$^{-2}$ s$^{-1}$. The total number of measurements used is about 215 000; the width of one
point in the distribution along the horizontal axis is $0.5\sigma$.
}\label{statsign}
\end{figure*}

An example of such a distribution for the averaging
time scale $\tau=50$~ ks is shown in Fig. 4 (dashed line).
For comparison, the solid line indicates a Gaussian
distribution that is slightly narrower (see below) than
the experimental one. To illustrate the quality of the
subtraction of the central source in the 511-keV line,
the dash-dotted line in Fig. 4 indicates the initial
distribution of flux significances. The total error estimated
by this method is valid only by assuming the
absence of any real flux spikes in the annihilation line
in the sky.

For the time scale under consideration, the total
error in units of the standard deviation related to
purely statistical fluctuations is $\sigma_{total}\simeq1.25$, which
corresponds to a systematic error $\sigma_{sys}\simeq0.75$
($\sigma^{2}_{total}=\sigma^{2}_{stat}+\sigma^{2}_{sys}$). 
 The contribution from the systematic error
increases with averaging time and reaches $\sigma_{sys}\simeq0.90$
 at $\tau=1000$~ ks. These systematic errors also
include the effects related to the possible presence of a
persistent annihilation emission source different from
the source at the Galactic center.

\begin{figure*}[t]
\centerline{\includegraphics[width=10cm,bb=24 275 515 680,clip]{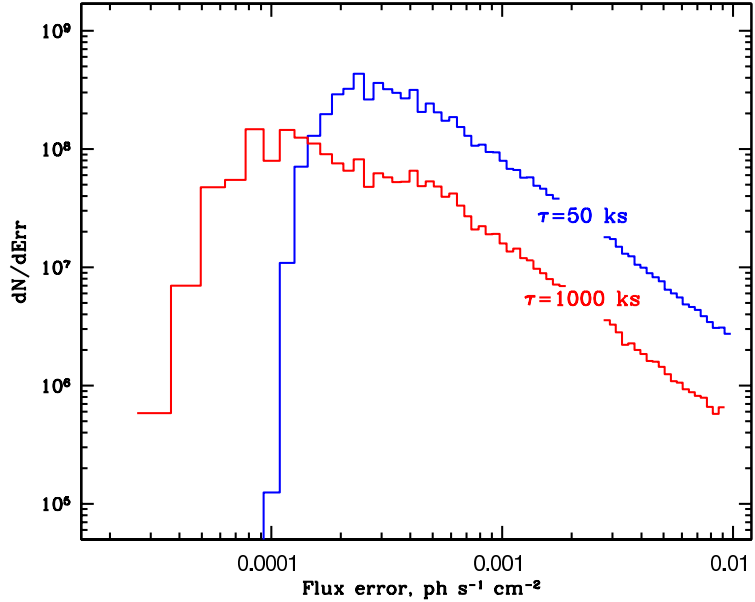}}

\vfill

\caption{Distribution of statistical flux measurements errors for all time bins over the sky map for two averaging time scales:
50 and 1000 ks. The errors are given in units of phot cm$^{-2}$ s$^{-1}$.
}\label{errdist}
\end{figure*}

Thus, given the systematic component of the error,
the actual significance of the detected outbursts is
lower by 25-35\%, depending on the averaging time
scale, but here we disregarded this error component.

\section*{RESULTS}

The effective area of the SPI spectrometer changes
over a wide range, depending on the source position
in its field of view. Accordingly, the same observation
will have greatly differing statistical errors in different
map pixels. Since our algorithm for analyzing the
light curves selects the most significant positive flux
deviations from the mean, statistical fluctuations can
lead us to the selection of a value with a high error,
while a slightly less significant point with a much
smaller statistical error will be rejected. To determine
more stringent upper limits on the flux variability in
the 511-keV line, we rejected all points with errors
above a certain value.

Thus, for example, to detect outbursts with a
flux of $1\times10^{-3}$ phot cm$^{-2}$ s$^{-1}$ 
 with a 5 $\sigma$ significance,
the statistical error should not exceed $2\times10^{-4}$
phot cm$^{-2}$ s$^{-1}$. For each time scale, we used
two limiting errors (Table 1).

\begin{table*}[t]
\noindent
\small
\centering
\caption{Limiting statistical errors of the experimental data
points used in our analysis for four different averaging time
scales.}\label{ttab_pred}
\centering
\vspace{1mm}
\begin{tabular}{c|c|c|c}
\hline\hline
Time scale, & Minimum  & Limiting & Limiting \\
ks       & error,& error 1, & error 2,\\
         & 10$^{-5}$ phot cm$^{-2}$ s$^{-1}$&
10$^{-4}$ phot cm$^{-2}$ s$^{-1}$ &10$^{-3}$ phot cm$^{-2}$ s$^{-1}$ \\
\hline
50   & 8 & 2 & 2 \\
100  & 8 & 2 & 2 \\
500  & 4 & 1 & 1 \\
1000 & 3 & 1 & 1 \\
\hline
\end{tabular}
\vspace{3mm}
\end{table*}

Figure 5 shows the distribution of statistical
errors in each of the time bins (for two averaging
time scales) over the entire sky map. The minimum
statistical flux measurement error for a time bin of
50 ks is $\sigma_{min}\sim8\times10^{-5}$
phot cm$^{-2}$ s$^{-1}$. This means
that we can detect an outburst with duration $t$ and
flux $5\times\sigma_{min} \times (5\times10^{4}/t)^{0.5}$
phot cm$^{-2}$ s$^{-1}$ with
a significance of 5 $\sigma$. The minimum error depends
on the averaging time scale and decreases to $3\times10^{-5}$
phot cm$^{-2}$ s$^{-1}$ for 1000-ks bins. However, as
we see from Fig. 5, most of the flux measurements
have a considerably larger statistical error, which
reduces the effective sensitivity of our analysis.

\begin{figure*}[t]
\centerline{\includegraphics[width=8cm]{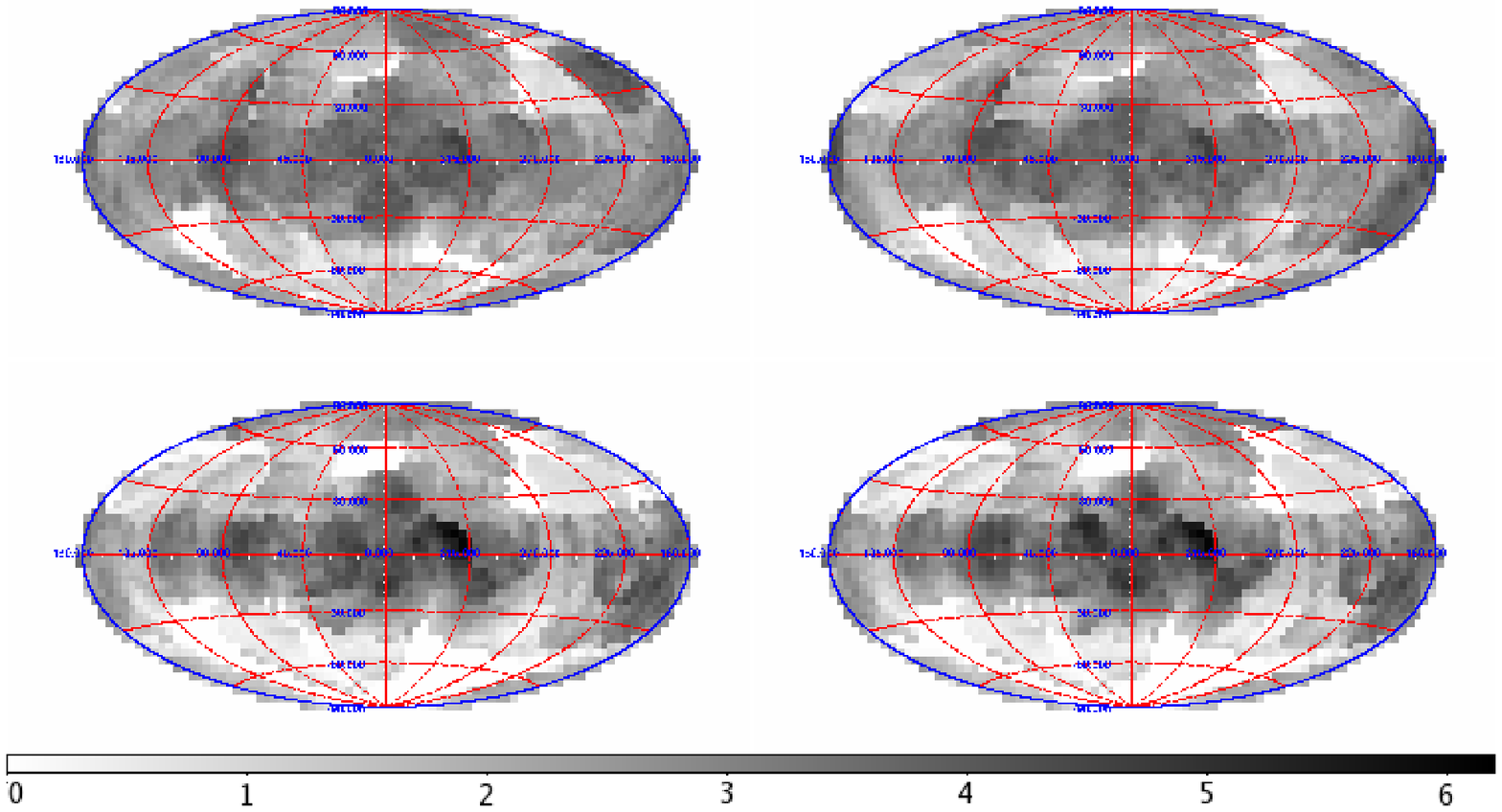}
\includegraphics[width=8cm]{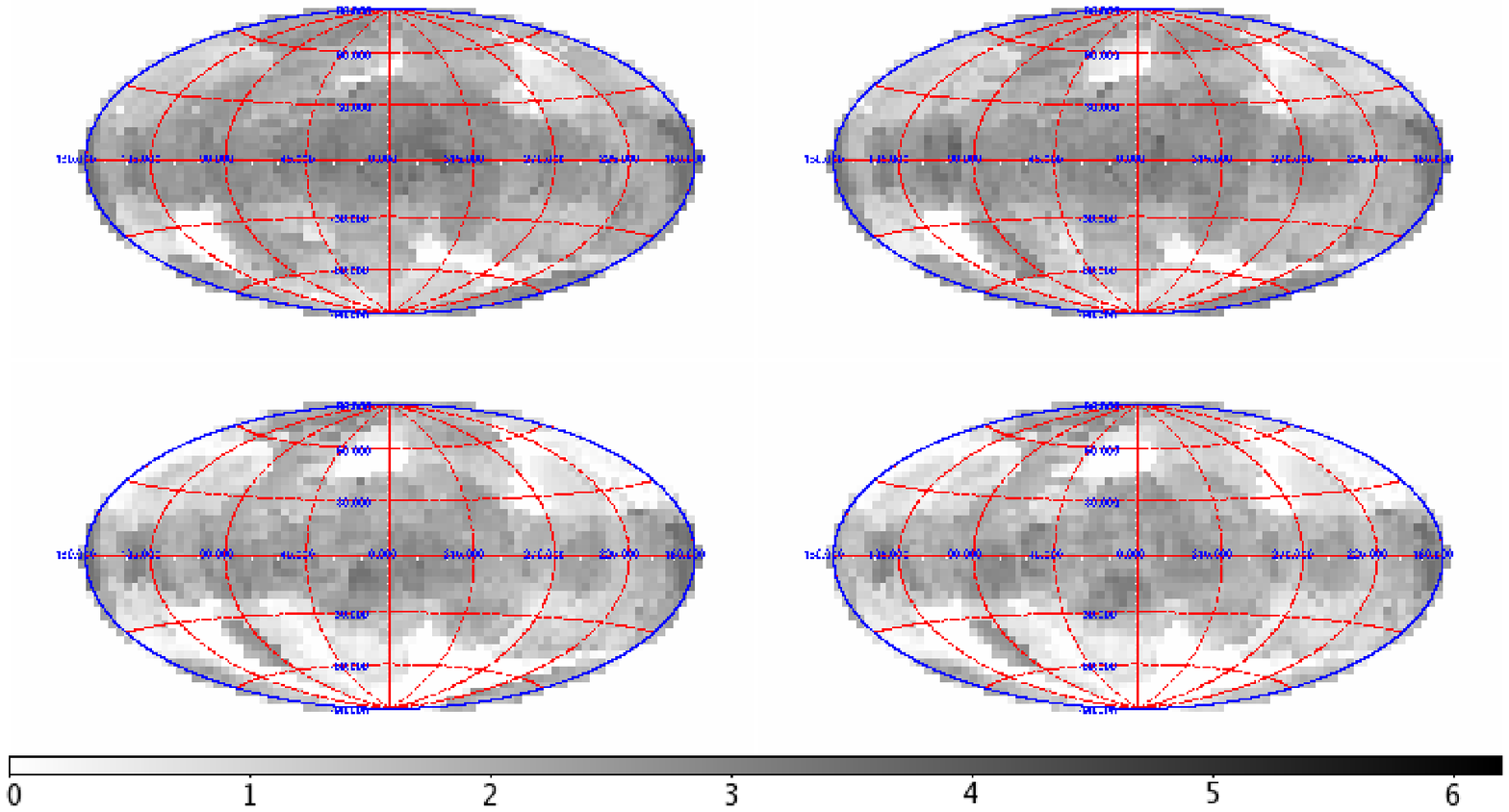}}

\vfill

\caption{(a) Map of maximum significance of a positive spike relative to the mean in the 508-514 keV energy band for all
averaging time scales (50, 100, 500, 1000 ks, from left to right) for the second set of limiting errors (Table 1); (b) similar map
obtained through Monte Carlo simulations (see the text). The images are
shown in units of the standard deviation.
}\label{sign}
\end{figure*}

As a result, we constructed the maps of the highest
significances of positive flux deviations in the
508-514 keV energy band from the mean in each
map pixel. Figure 6a presents our results for all averaging
time scales using the second set of limiting
errors (see Table 1). We see that the positions of the
most significant spikes in flux in the 511-keV line
coincide with the region around the Galactic center
and regions with a long observing time.

The increase in significance in these regions can
be related to two effects: a large number of independent
measurements (see below) and a possible
contribution from nonideal subtraction of the diffuse
source at the Galactic center.

To estimate the effect of an increase in the significance
of the flux spike in the 511-keV line in the
sky regions where the number of independent flux
measurements is great, Fig. 7 shows the distribution
of the probabilities to detect a maximum spike
with a significance exceeding a given one by chance
for three different numbers of independent trials. For
high significance levels, this probability is, obviously,
equivalent to the probability to detect such a spike for
a Gaussian distribution multiplied by the number of
independent trials.

The number of independent time bins in our analysis
depends on the averaging time scale, the area
of the analyzed sky region, and the maximum error
in this averaging. The total number of INTEGRAL
pointings used here is about 35 000. The number of
independent points decreases as the averaged light
curves are constructed.

\begin{figure*}[t]
\centerline{\includegraphics[width=10cm,bb=85 250 511 671,clip]{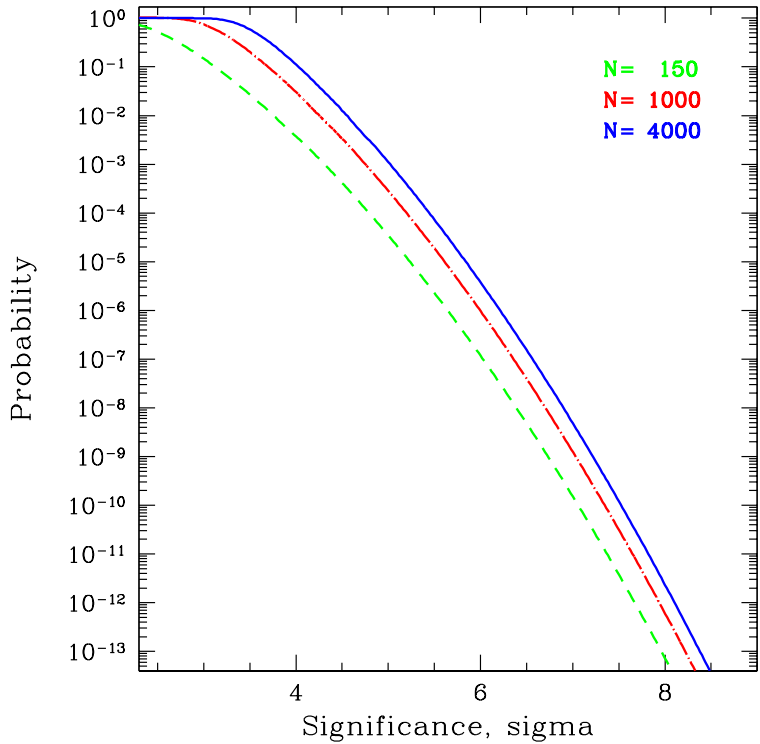}}

\vfill

\caption{Distribution of the probabilities to detect a random spike
  with a maximum significance from the entire data set exceeding
a given one. The dotted, dash-dotted, and solid lines correspond to
the numbers of independent trials $N=150$, $N=1000$, and $N=4000$, respectively.
}\label{probab}
\end{figure*}

For example, the maximum recorded significance
in the entire sky is $\sim$$4.6\sigma$
 for 100-ks time bins with
errors of less than $2\times10^{-4}$
phot cm$^{-2}$ s$^{-1}$. If the
number of independent bins there is $\sim$2500, then the
probability to find a spike with such a significance
by chance is $P\sim5\times10^{-3}$. Although the probability
to detect such a spike only through statistical
fluctuations is low, given the remaining inaccuracies
in modeling the background (fractions of the
statistical error), we use the corresponding value of
$P\sim5\times10^{-3}$
phot cm$^{-2}$ s$^{-1}$
 (Table 2) as a conservative
upper limit on the maximum flux in an outburst of
annihilation radiation.

For clarity, Fig. 6b shows maps of the maximum
significances of spikes similar to those shown in
Fig. 6a but obtained by the Monte Carlo method:
instead of the actually measured count rates in the
508-514 keV energy band, we substituted quantities
with a Gaussian distribution with a mean of 0 and a
variance equal to the error in a given measurement
into the algorithm of our analysis. We see that
purely statistical fluctuations (without any systematic
effects) can lead to expected spikes comparable to the
observed ones. For example, the Monte Carlo method
shows that the expected maximum flux deviation from
the mean for an averaging time scale of 100 ks and
the second set of limiting errors (Table 1) at the
Galactic center is $3\times10^{-3}$
phot cm$^{-2}$ s$^{-1}$ with a
significance of $\sim$$3.5\sigma$. In the actual data, these values
are $1\times10^{-3}$ phot cm$^{-2}$ s$^{-1}$
 and 4.2$\sigma$, respectively.

Using the total observing time of a particular sky
region, we can obtain upper limits on the rate of
outbursts with a given duration and flux. Such estimates
were made for a confidence level of $10^{-3}$ by
assuming three different scenarios for the distribution
of outbursts in the sky (see Fig. 1):

(i) uniform near the Galactic center (the distance
from the center is R$<$25\deg);

(ii) uniform in the Galactic plane 
(-120\deg$<l<$120\deg, -25\deg$<b<$25\deg);

(iii) uniform over the entire sky.

The final results of our analysis are contained in
Table 2: the maximum recorded significance ($S_{max}$,
in units of the standard deviation) of an outburst in
a given region and the corresponding flux deviation
from the mean ($F_{max}$, in units of phot
cm$^{-2}$ s$^{-1}$) are
given for each of the sky regions under consideration
(see above). All of the results are presented for
four averaging time scales and two levels of limiting
errors (Table 1). The value of $\nu$ limits the rate of
outbursts with a flux $F>F_{max}$ in a given sky region
at a confidence level of $10^{-3}$
 (in units of yr$^{-1}$). The
calculations were performed by assuming an equal
probability of outbursts at any location of the region
under consideration.

\section*{DISCUSSION}

The maximum significance of the flux variability
in the 511-keV line is observed near the Galactic
plane at a Galactic latitude of $-30^{\circ}$. This is the most
interesting feature on the map, since the low-mass
black-hole binary GRO J1655-40 can be a potential
source of this outburst. The observed outburst in the
508-514 keV energy band coincides in time with an
intense outburst in the standard X-ray energy band
(Markwardt and Swank 2005) detected by many observatories
in 2005.

Unfortunately, the duration of the source's observations
is too short to analyze this event in detail. The
observed flux excess above the mean recorded by SPI
in the 508-514 keV energy band is $\sim$$4\times10^{-4}$
 and $\sim$$2\times10^{-3}$ phot cm$^{-2}$ s$^{-1}$
 for the first and second
sets of limiting errors, respectively. However, the absolute
significance of this result (Table 2) is too low to
talk about a reliable detection of the outburst.

\begin{figure*}[t]
\centerline{\includegraphics[width=10cm,bb=15 275 515 680,clip]{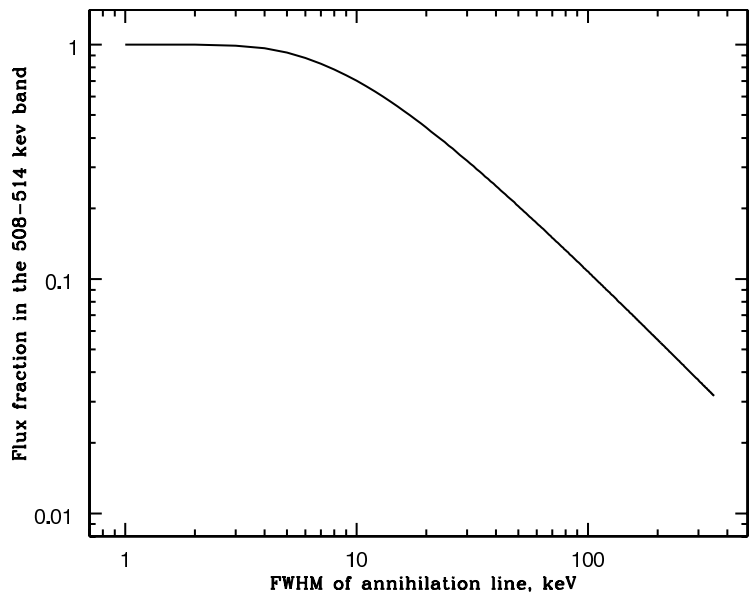}}

\vfill

\caption{Flux fraction in the line centered at 511 keV falling
within the 508-514 keV energy band versus FWHM of
this line.}\label{flpart}
\end{figure*}

It is important to note that the upper limits given
in Table 2 refer only to the narrow emission line.
If compact objects are the sources of the emission
in the annihilation line, then its width can be much
larger than several keV; it can also be shifted as a
whole. In our case, the line flux will not be entirely
in the 508-514 keV energy band and the algorithm
we use will not show its significant variations. In
Fig. 8, the fraction of the flux in the line centered at
511 keV falling within the 508-514 keV energy band
is plotted against the FWHM of this line. Thus, a
flux in the 508-514 keV energy band of only 
$\sim$$2\times10^{-4}$ phot cm$^{-2}$ s$^{-1}$
 corresponds to a line flux of
$\sim$$6\times10^{-3}$ phot cm$^{-2}$ s$^{-1}$
 from Nova Musca at a 58-keV width of the line centered at 476 keV (Sunyaev
et al. 1992; Gilfanov et al. 1991). Such an insignificant
change in flux cannot be recorded with a significance
of more than $\sim$$2\sigma$ at an outburst duration
of 100 ks even if the minimum statistical error is
achieved for the time scale in question (Table 1).

Our upper limit on the flux variability in the
narrow 511-keV line from 1E 1740.7-2942 is $\sim$$5\times10^{-4}$
phot cm$^{-2}$ s$^{-1}$
 for an averaging time scale of
100 ks and the first set of limiting errors. Bouchet
et al. (2009) present the results of a detailed spectral
analysis of the emission from this microquasar based
on three-year-long SPI and IBIS observations. The
authors found neither a persistent, nor transient (on
a time scale of 0.5--1 day), nor broad, nor narrow
emission line near 511 keV in the source's spectrum.
They provide upper limits on the line detection on a
time scale of one day, $\sim$$7\times10^{-4}$
phot cm$^{-2}$ s$^{-1}$.

\section*{CONCLUSIONS}

We presented the results of a systematic search
for outbursts in the narrow positron annihilation line
on various time scales ($5\times10^{4}$~ -- ~$10^{6}$~ s) based on the
SPI/INTEGRAL data obtained from 2003 to 2008.

We showed that the background radiation level
could be predicted on the basis of a reference energy
band. Here, the 520-600 keV energy band containing
no contribution from the narrow 511-keV line and
the ortho-positron continuum was used for this purpose.
A serious advantage of the background model
construction based on a reference energy channel is
an effective subtraction of the systematic features that
manifest themselves in both bands.

We showed that no outbursts were observed with
a statistical significance higher than $\sim$$6\sigma$
 for any
of the time scales considered in the entire sky. The
maximum annihilation flux amplitude depends on
the maximum errors and time scale used (Table 2).
We estimated the effect of statistical fluctuations on
the detection of outbursts of annihilation radiation.
Thus, for example, using the Monte Carlo method,
we showed that a maximum flux deviation from the
mean of $3\times10^{-3}$ phot cm$^{-2}$ s$^{-1}$
was reached at the
Galactic center with a significance of $\sim$$3.5\sigma$ for an
averaging time scale of 100 ks and statistical errors
no larger than $2\times10^{-3}$ phot cm$^{-2}$ s$^{-1}$. 
In contrast,
the experimental value of the most significant positive
flux deviation from the mean in this region is $1\times10^{-3}$
phot cm$^{-2}$ s$^{-1}$ with a significance of $\sim$$4.2\sigma$.

The statistical flux measurements errors in the
508-514 keV energy band allow an outburst of duration
$t$ and flux $5\times\sigma_{min} \times (5\times10^{4}/t)^{0.5}$
phot cm$^{-2}$ s$^{-1}$
(where $\sigma_{min}\sim8\times10^{-5}$
 phot cm$^{-2}$ s$^{-1}$) to be detected 
with a significance of 5$\sigma$.

Based on the exposure achieved in $\sim$6 yr of INTEGRAL
operation, we provide conservative upper
limits on the rate of outbursts with a given duration
and flux in different parts of the sky (for a confidence
level of $10^{-3}$). Thus, for example, one might expect no
more than $\sim$3 outbursts per year in the entire sky.

Based on the recorded variability, we provide constraints
on the flux in the narrow annihilation emission
line in the spectrum of the transient GRO J1655-40 
during its outburst in 2005. The most conservative
upper limit does not exceed $\sim$$2\times10^{-3}$
phot cm$^{-2}$ s$^{-1}$.

\centerline{}
\centerline{}
\centerline{}
\centerline{}

{\bf ACKNOWLEDGMENTS}

We wish to thank R.A. Sunyaev for the discussion
of our results and helpful remarks. 
 This work was supported
by grant no. NSh-5579.2008.2 from the President
of Russia, the ``Origin, Structure, and Evolution of Objects
in the Universe'' Program of the Presidium of the Russian
Academy of Sciences, and the Russian Foundation
for Basic Research (project no. 07-02-01051). We
used data from the INTEGRAL Science Data Center
(Versoix, Switzerland) and the Russian INTEGRAL
Science Data Center (Moscow, Russia).

\pagebreak

\begin{landscape}
\begin{table*}[h]
\noindent
\small
\centering
\caption{Results of our analysis. For each of the averaging time scales under consideration, we give the maximum recorded significance of an outburst in a
specific sky region ($S_{max}$, in units of the standard deviation) and the corresponding flux deviation from the mean ($F_{max}$, in units of
phot cm$^{-2}$ s$^{-1}$). The second column provides the rate of outbursts in a given sky region ($\nu$, in units of yr$^{-1}$).}\label{tab_res}
\centering
\vspace{1mm}
\begin{tabular}{c|c|c|c|c|c|c|c|c|c}
\hline\hline
Sky region  & Rate of &  \multicolumn{8}{c}{Averaging time scale, ks} \\
\cline{3-10}
        & outbursts, $\nu$ &  \multicolumn{2}{c|}{50}& \multicolumn{2}{c|}{100}&\multicolumn{2}{c|}{500}& \multicolumn{2}{c}{1000}  \\
\cline{3-10}
  & $\nu$ &$S_{max}$&$F_{max}$&$S_{max}$&$F_{max}$&$S_{max}$&$F_{max}$&$S_{max}$&$F_{max}$\\
\hline
\multicolumn{10}{c}{Limiting errors 1} \\
\hline
Galactic center & 10.1 & 3.9 & $6.9\times10^{-4}$ & 4.1 & $4.2\times10^{-4}$ & 4.9 & $3.5\times10^{-4}$ & 4.9 & $3.4\times10^{-4}$ \\
Galactic plane & 3.8  & 4.6 & $8.7\times10^{-4}$ & 4.5 & $8.9\times10^{-4}$ & 5.0 & $3.5\times10^{-4}$ & 4.9 & $3.8\times10^{-4}$ \\ 
Entire sky & 2.9  & 4.7 & $8.8\times10^{-4}$ & 4.6 & $4.1\times10^{-4}$ & 5.0 & $3.5\times10^{-4}$ & 4.9 & $3.8\times10^{-4}$ \\
\hline
\multicolumn{10}{c}{Limiting errors 2} \\
\hline
Galactic center & 10.1 & 4.3 & $2.1\times10^{-3}$ & 4.2 & $1.0\times10^{-3}$ & 5.3 & $9.7\times10^{-4}$ & 5.3 & $9.7\times10^{-4}$ \\
Galactic plane & 3.8  & 4.7 & $1.9\times10^{-3}$ & 4.7 & $1.8\times10^{-3}$ & 6.1 & $1.7\times10^{-3}$ & 6.1 & $1.7\times10^{-3}$ \\
Entire sky & 2.9  & 4.7 & $1.9\times10^{-3}$ & 4.7 & $1.8\times10^{-3}$ & 6.1 & $1.7\times10^{-3}$ & 6.1 & $1.7\times10^{-3}$ \\
\hline
\end{tabular} 
\vspace{3mm}
\end{table*}
\end{landscape}
\clearpage

{\bf REFERENCES}

1. R. M. Bandyopadhyay, J. Silk, J. E. Taylor, et al.,
MNRAS 392, 1115 (2009).

2. L. Bouchet, P.Mandrou, J. P.Roques, et al., Astrophys.
J. 383, L45 (1991).

3. L. Bouchet, M. Del Santo, E. Jourdain, et al., Astrophys.
J. 693, 1871 (2009).

4. E. Churazov, R. Sunyaev, S. Sazonov, et al., MNRAS 357, 1377 (2005).

5. E. Churazov, S. Sazonov, S. Tsygankov, et al., (2010,
submitted).

6. M. Forman, R. Ramaty, and E. Zweibel, in Physics of
the Sun, Vol. II, Ed. by P.Sturrok (1986), p. 249.

7. M.R.Gilfanov, R. A. Sunyaev,E. M. Churazov, et al.,
Pis'ma Astron. Zh. 17, 1059 (1991) [Sov. Astron.
Lett. 17, 437 (1991)].

8. M. Gilfanov, E. Churazov, R. Sunyaev, et al., Astrophys.
J. Suppl.Ser. 92, 411 (1994).

9. A. Goldwurm, J. Ballet, B. Cordier, et al., Astrophys.
J. 389, L79 (1992).

10. W. N. Johnson, III, F. R. Harnden, Jr., R. C. Haymes,
et al., Astrophys. J. 172, L1 (1972).

11. J. Knodlseder, P.Jean, V. Lonjou, et al., Astron. Astrophys.
441, 513 (2005).

12. C. B. Markwardt and J. H. Swank, Astron. Telegram
414 (2005).

13. N. Prantzos, in Proc. of the 5th INTEGRAL Workshop
on the INTEGRAL Universe,Ed. byV. Schoenfelder,
G. Lichti, and C. Winkler (ESA SP-552,
2004), p. 15.

14. W. R. Purcell, L.-X. Cheng, D. D. Dixon, et al., Astrophys.
J. 491, 725 (1997).

15. G. R. Riegler, J. C. Ling, W. A. Mahoney, et al.,
Astrophys. J. 248, L13 (1981).

16. S. J. Sturner, C. R. Shrader, G. Weidenspointner,
et al., Astron. Astrophys. 411, L81 (2003).

17. R. Sunyaev, E. Churazov, M. Gilfanov, et al., Astrophys.
J. 383, L49 (1991).

18. R. Sunyaev, E. Churazov, M. Gilfanov, et al., Astrophys.
J. 389, L75 (1992).

19. B. J. Teegarden, Astrophys. J. Suppl. Ser. 92, 363
(1994).

20. B. J. Teegarden and K.Watanabe, Astrophys. J. 646,
965 (2006).

21. G. Vedrenne, J.-P. Roques, V. Schonfelder, et al.,
Astron. Astrophys. 411, L63 (2003).

22. G. Weidenspointner, G. Skinner, P. Jean, et al., Nature
451, 159 (2008).

23. C. Winkler, T. J.-L. Courvoisier, G. Di Cocco, et al.,
Astron. Astrophys. 411, L1 (2003).

\hspace{90 mm} Translated by V. Astakhov

\end{document}